\documentclass[12pt]{article}
\usepackage{graphicx}
\begin{document}
\centerline{\bf \Large Applications and Sexual Version of }
\centerline{\bf \Large a Simple Model for Biological Ageing}
\bigskip

\centerline{A. O. SOUSA$^1$, S. MOSS DE OLIVEIRA$^1$ and D.STAUFFER$^2$}
\bigskip

\noindent $^1$Instituto de F\'{\i}sica, Universidade Federal Fluminense,
Av. Litor\^anea s/n, Boa Viagem, 24210-340 Niter\'oi, RJ, Brasil.

\noindent $^2$Inst. for Theor. Physics, Cologne University, 
D-50923 K\"oln, Euroland.

\medskip
e-mail: sousa@if.uff.br, suzana@if.uff.br, stauffer@thp.uni-koeln.de

\bigskip

{\bf Abstract} We use a simple model for biological ageing to study  
the mortality of the population, obtaining a good agreement with the 
Gompertz law. We also simulate the same model on a square lattice, 
considering different strategies of parental care. The results are in 
agreement with those obtained earlier with the more complicated Penna 
model for biological ageing. Finally, we present the sexual version 
of this simple model.


\section{Introduction}

One theory for biological ageing is the accumulation of bad inherited  
mutations. While the well known Penna model \cite{penna} uses a 
genome represented by a bitstring, this simple model \cite{dresden}
reduces the genetic information, transmitted with mutations from one 
generation to the next, to two integers, the minimum reproduction age $a_m$
and the genetic death age $a_d$. At each iteration, every individual with age
$a$ between $a_m$ and $a_d$ gets at most one child (birth rate $b =
1$) which inherits these numbers,
apart from random changes (mutations) by $\pm 1$. Individuals die with certainty
if their age reaches $a_d$, and they die earlier with probability $N(t)/N_{max}$
at every iteration, where $N(t)$ is the actual population and $N_{max}$ a
so-called carrying capacity taking into account the limitations of food and
space (Verhulst factor). The probability $p_b(i)$ for individual $i$
to get a child takes into account
the well known (see e.g. \cite{tradeoff}) trade-off between fecundity and 
survival and is the lower the larger the difference $a_d-a_m$ is:
$$ p_b(i) = (1 + 0.08)/(a_d(i)-a_m(i) + 0.08)$$
where the constant 0.08 ensures that the denominator is never zero.
(For the mutations we require $ 0 \le a_m < a_d \le 32$ to facilitate comparison
with the Penna model which typically has 32 age intervals.)

This simple model allows for a self-organization (``emergence'') of a broad
distribution of $a_m$ values, and a relatively more narrow distribution of
$a_d$ values \cite{dresden} (similar but not identical to those shown 
below in Fig.6; the $a_m$ distribution gets more narrow relatively if, as in 
section 4, an increase of $a_m$ is coupled to an increase of $a_d$.). 
It explained the catastrophic senescence of Pacific
Salmon \cite{ortmanns}, the vanishing of cod in the northwest Atlantic through 
overfishing \cite{radomski}, and allowed to take into account the social needs 
of humans for a minimum population size \cite{radomski}. But only with a birth
rate increasing in old age \cite{makowiec} could a reasonable agreement with 
the Gompertz law be achieved.

\begin{figure}[H]
\begin{center}
\includegraphics[angle=-90,scale=0.5]{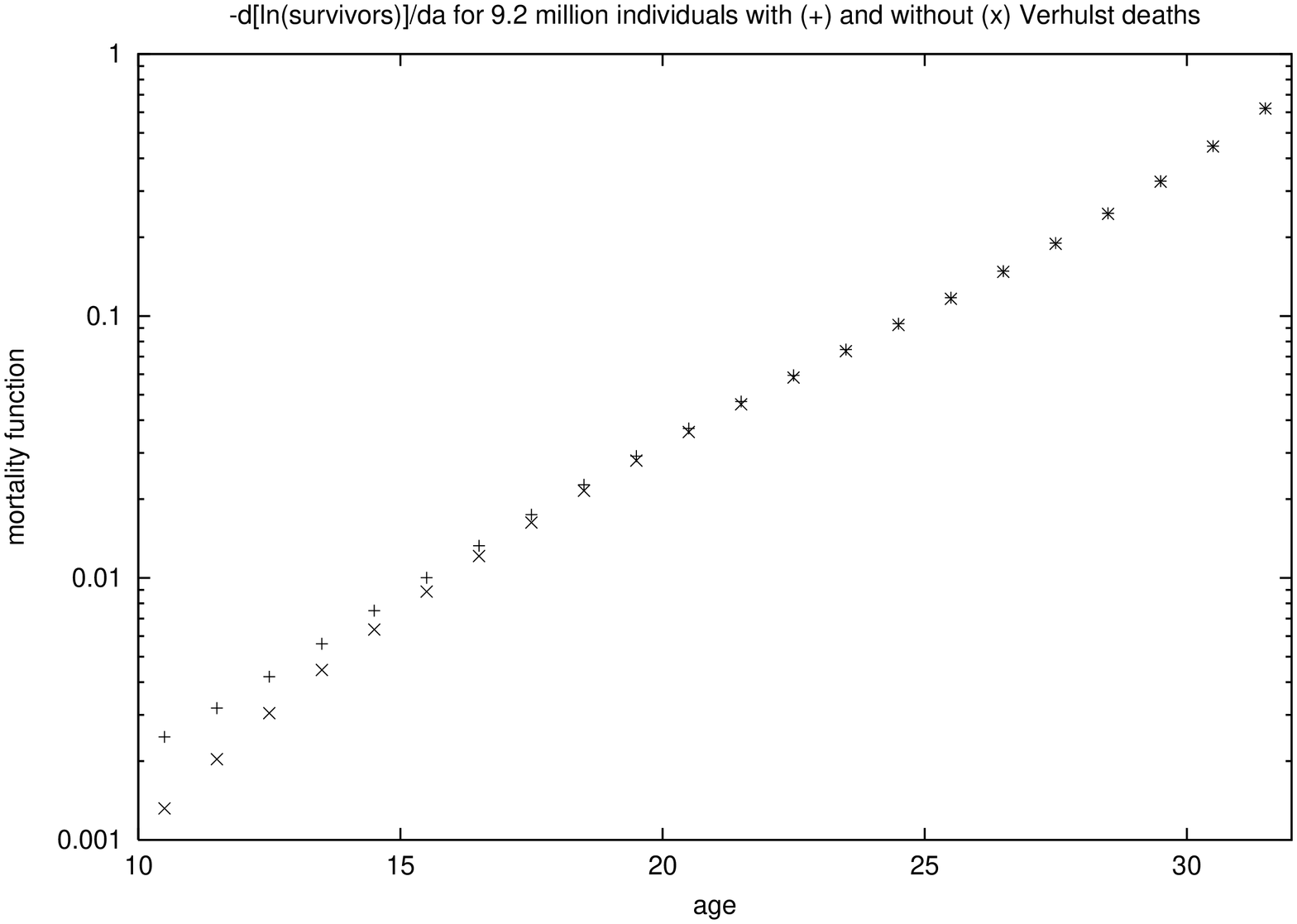}
\end{center}
\caption{
Mortality function $\mu$ versus age (in iterations) giving rough agreement 
with the Gompertz law; $\beta = 0.7$ with and without Verhulst deaths
using 9 million individuals.
}
\end{figure}    

In the next section we present the mortality obtained with a slightly modified
version of this model, which then agrees with
the Gompertz law. In section 3 we present the results of the model 
when the individuals are placed on a square lattice and live under
some parental care strategies. Finally, in section 4 we present the 
sexual version of the model and conclusions.

\section{The Gompertz law}
To get the Gompertz law of an exponentially increasing mortality function
of age $a$, we now follow \cite{thoms} and assume the genetic deaths to be 
probabilistic instead of deterministic, with probability 
$p_d = 1/\{1 + \exp[\beta(a_d-a)]\}$ 
and $\beta = 0.7$; the deterministic case is
recovered for $\beta \rightarrow \infty$. The survival probability thus
decays like a Fermi function in physics. Fig.1 shows a good agreement with
the exponential increase of the mortality function \cite{nusbaum,book}
$\mu(a) = \ln[S(a-1/2)/S(a+1/2)]$
for humans at middle age (where $S(a)$ is the number of survivors at age $a$;
$\mu$ approximates the derivative $-d(\ln S)/da$). For the deterministic case,
$\beta \rightarrow \infty$, the analogous plot is strongly curved. Alternatively
one may omit \cite{pmco} the condition that no age exceeds 32, increase the
constant from 0.08 to about 2, and set $\beta = 0.3$; then Fig.2 shows a similar
increase of the mortality function with age, with a reasonable 
distribution of genetic death ages which is no longer cut off at 32.

\begin{figure}[H]
\begin{center}
\includegraphics[angle=-90,scale=0.5]{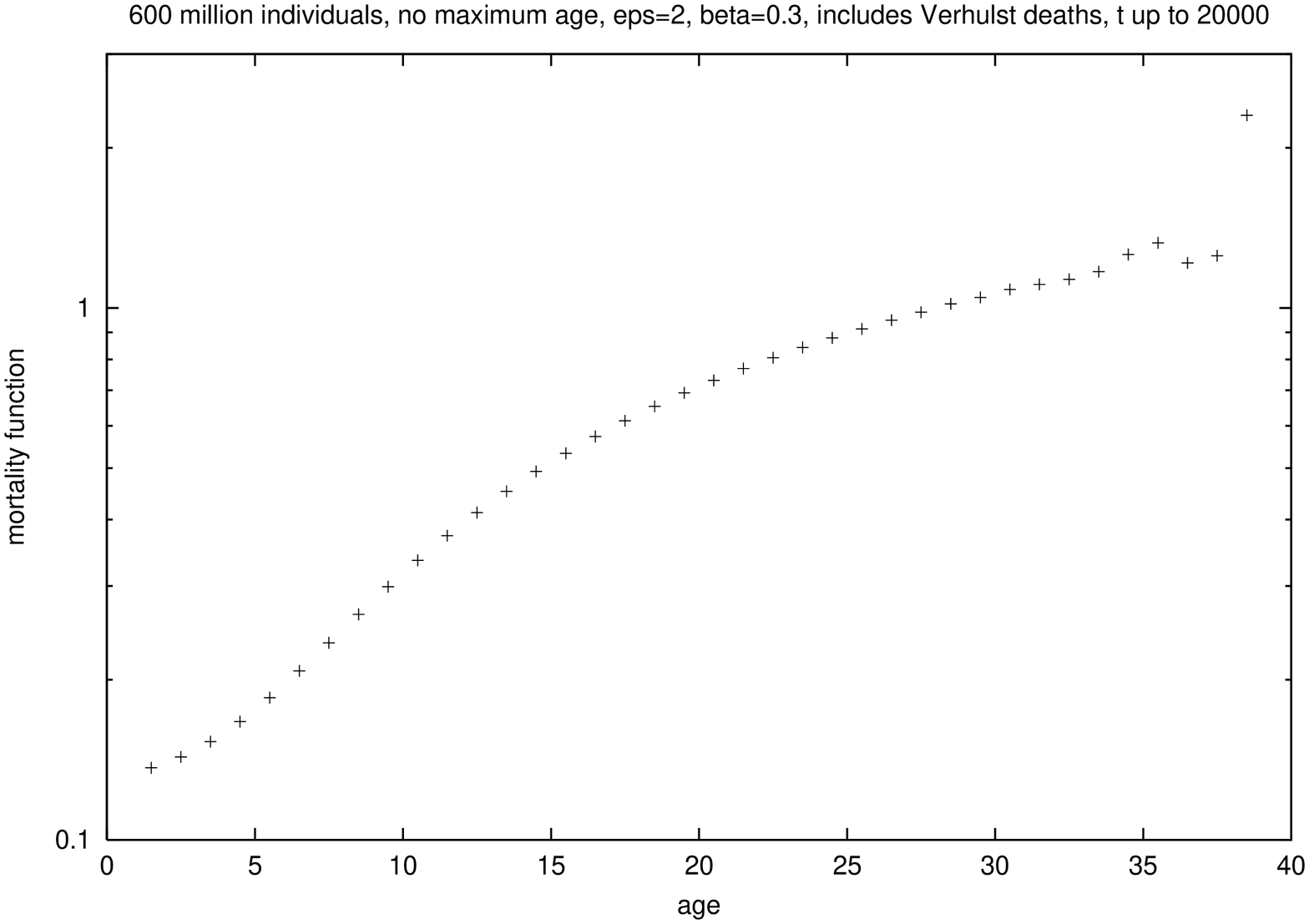}
\end{center}
\caption{
As Fig.1 but $\beta = 0.3$ with 609 million individuals including Verhulst 
deaths, no maximum age and 0.08 replaced by 2.0 in the birth rate.
}
\end{figure}    

\section{Parental care on a square lattice}

\begin{figure}[H]
\begin{center}
\includegraphics[angle=-90,scale=0.50]{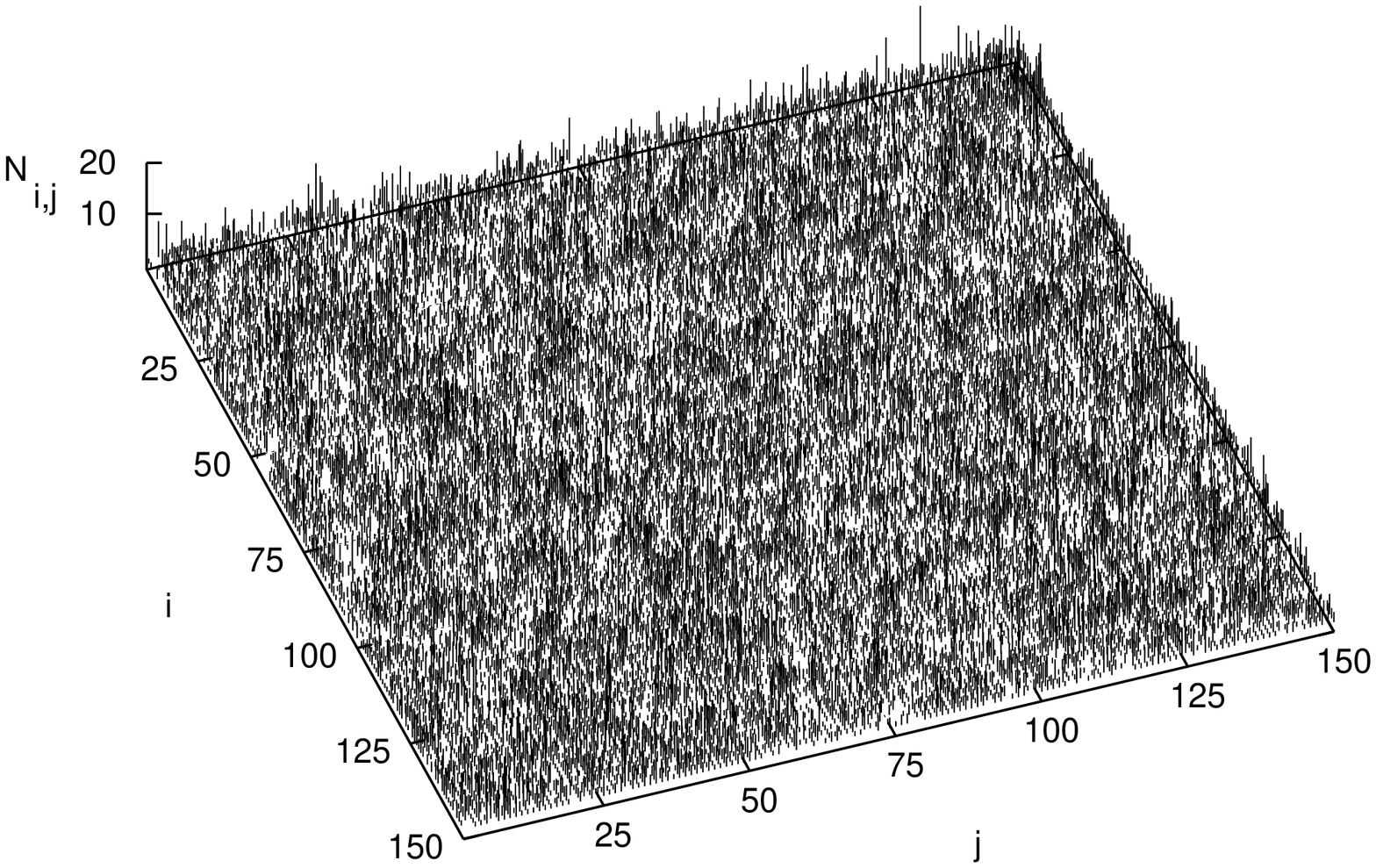}
\end{center}
\caption{
Final distribution of the population on the lattice considering
that if the mother moves, she brings the children under maternal
care (age $\le2$) with her (case a). 
The final distribution of the lattice considering
that the mother cannot move, if she has children under maternal (case b),
looks very similar (not shown). 
}
\end{figure} 

Now we return to the original deterministic version of the model and 
place the individuals on a square lattice. The carrying capacity 
$N_{max}$ is now replaced by a maximum occupation per site
$N_{max}(i,j)$, such that the new Verhulst probability for an 
individual to die due to environmental restrictions depends on the 
current number of individuals $N_{(i,j)}(t)$ on each site. We give to 
each individual a probability $p_m$ to 
move to the neighbouring site that presents the smallest occupation, 
if this occupation is also smaller or equal to that of the current 
individual's site. The strategy of child-care consists in defining 
a maternal care period $A_{pc}$ during which no child can move. 
We considered the following conditions: (a) if the mother moves, 
she brings the young children with her; (b) the mother cannot move if 
she has any child still under maternal care. In Fig. 3  we 
show the lattice configurations after 800,000 steps for case (a). 
The result of case (b) is very similar and not shown. These same
strategies were simulated before using the 
Penna model \cite{lattice}, giving these same results. Also following 
\cite{lattice}, we simulated a third case which corresponds to case 
(a) with the restriction that if the mother dies, her children still 
under maternal care have a further probability to die. Figure 4 shows 
the spatial configuration 
for this case (c). Again the result is the same obtained with the 
Penna model, despite of the simplicity of the present model. Fig.5 compares
the population sizes of the three cases. The parameters used to
produce figures 3, 4 and 5 are: initial population = 10,000
individuals; maximum occupation per site = 30 individuals; initial
minimum reproduction age $a_m = 1$; initial genetic death age $a_d =
16$; birth rate $b = 2$; maternal care period $A_{pc} = 2$ and lattice
size = $150 \times 150$.

\begin{figure}[H]
\begin{center}
\includegraphics[angle=-90,scale=0.50]{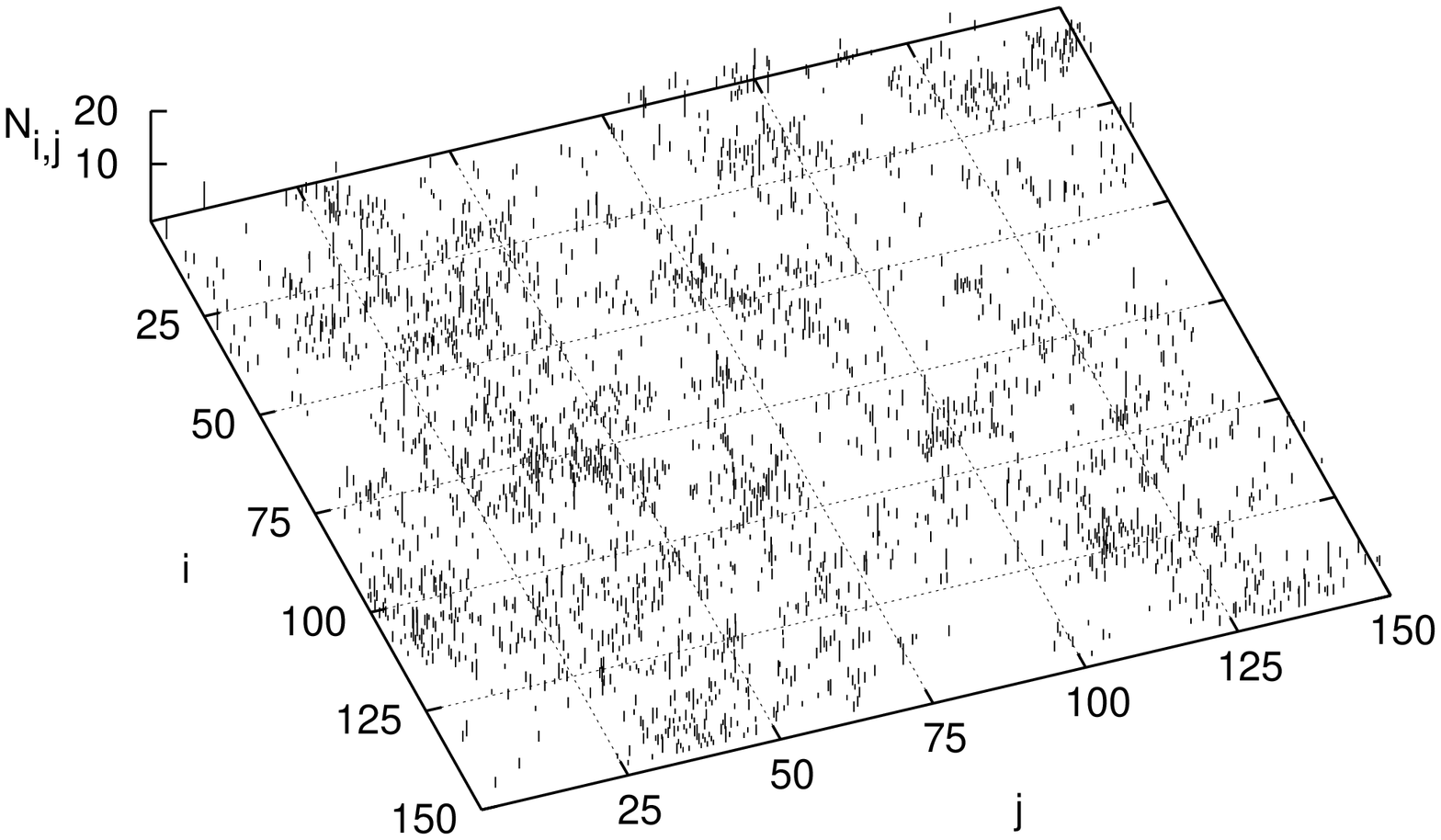}
\end{center}
\caption{
Final distribution of the lattice considering that if the mother moves she
brings the children under maternal care with her, and that if she dies,
her children have a probability $0.9$ to die in their first year of life,
and $0.3$ in their second year of life (case c). In this case we used
a maximum occupation per site equal to 42 individuals. Even so, the final
configuration of the lattice is much smaller than those obtained in
cases (a) and (b).} 
\end{figure}  

\begin{figure}[H]
\begin{center}
\includegraphics[angle=-90,scale=0.5]{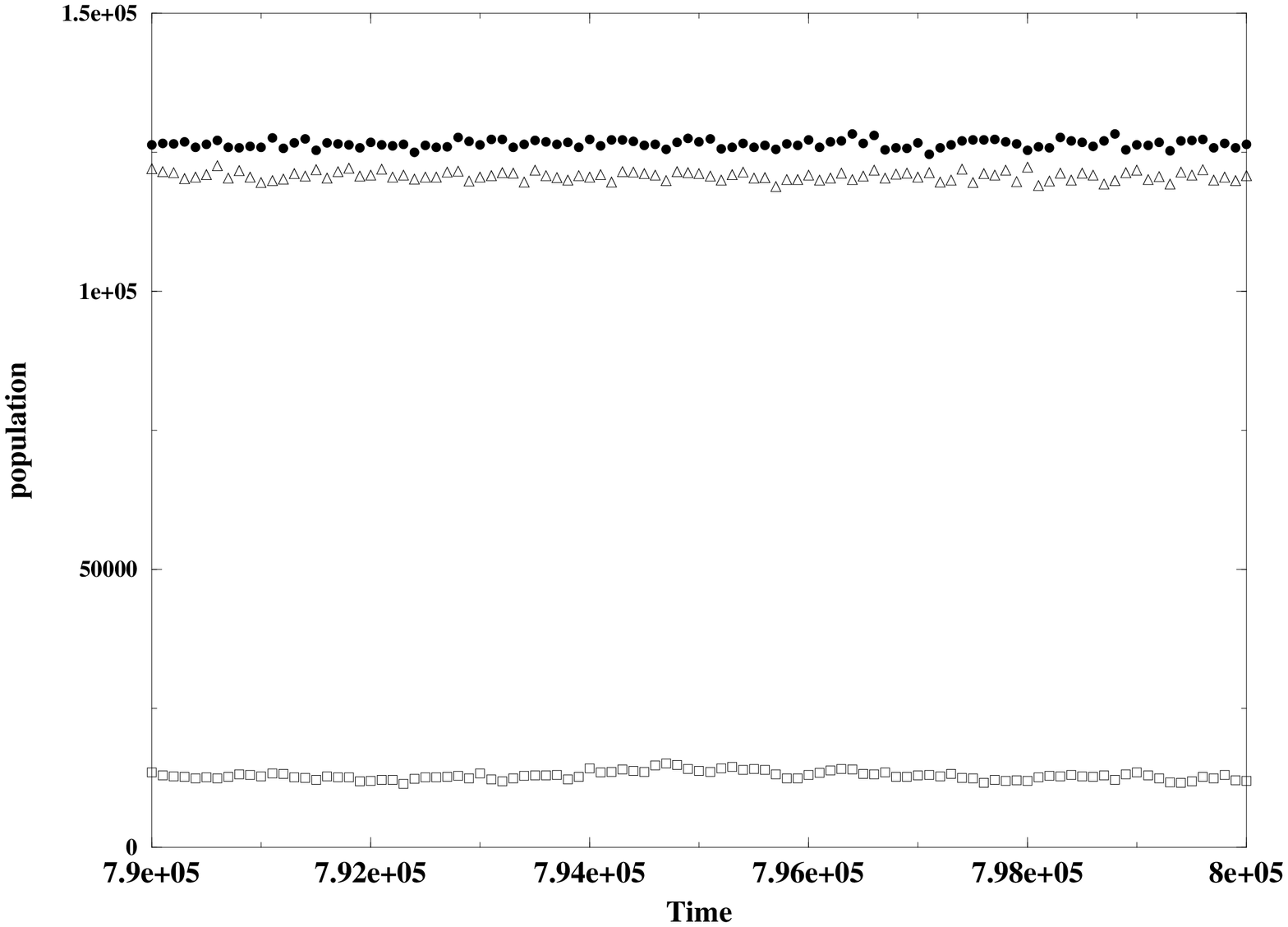}
\end{center}
\caption{
Number of individuals as a function of time for cases: a (circles),
b (triangles) and c (squares).
}
\end{figure}

\section{The sexual version} 

Now the population consists of males and females. At each iteration,
every female with age $a$ between $a_m$ and $a_d$ randomly chooses a
male, also with age between $a_m$ and $a_d$, to mate. The child
randomly inherits the $a_m$ and $a_d$ values from one of the parents,
independently. Mutations are introduced in the following way: two
random numbers are generated; if they are both positive, $a_m$ and
$a_d$ are both increased by one. If they are both negative, $a_m$ and
$a_d$ are decreased by one; if the signs of the random numbers are
different, no mutation occurs. Notice that in this way the difference
$a_m - a_d$ is maintained constant. This strategy simulates the fact
that a recessive mutation, to be effective, must be inherited from
both parents; alternatively, it can be interpreted as antagonistic
pleiotropy, the trade-off between fecundity and longevity
\cite{book,tradeoff}.
One may also regard this sexual version as an alternative to the requirement
of a minimal population for human society \cite{radomski}: a single individual
cannot reproduce.
In Fig. 6 we show the histograms of $a_m$ and
$a_d$, that is, the number of individuals with a given value of $a_d$
and the number of individuals with a given value of $a_m$. It can be
seen that there is also, as in the asexual case \cite{dresden}, a
self-organization of distributions of $a_m$ and $a_d$.  We have also
tried to introduce the mutations in $a_m$ and $a_d$ independently of
each other: Then no self-organization appears, that is, the $a_m$
value goes always to 1 and the $a_d$ value, to 32.

\begin{figure}[H]
\begin{center}
\includegraphics[angle=-90,scale=0.5]{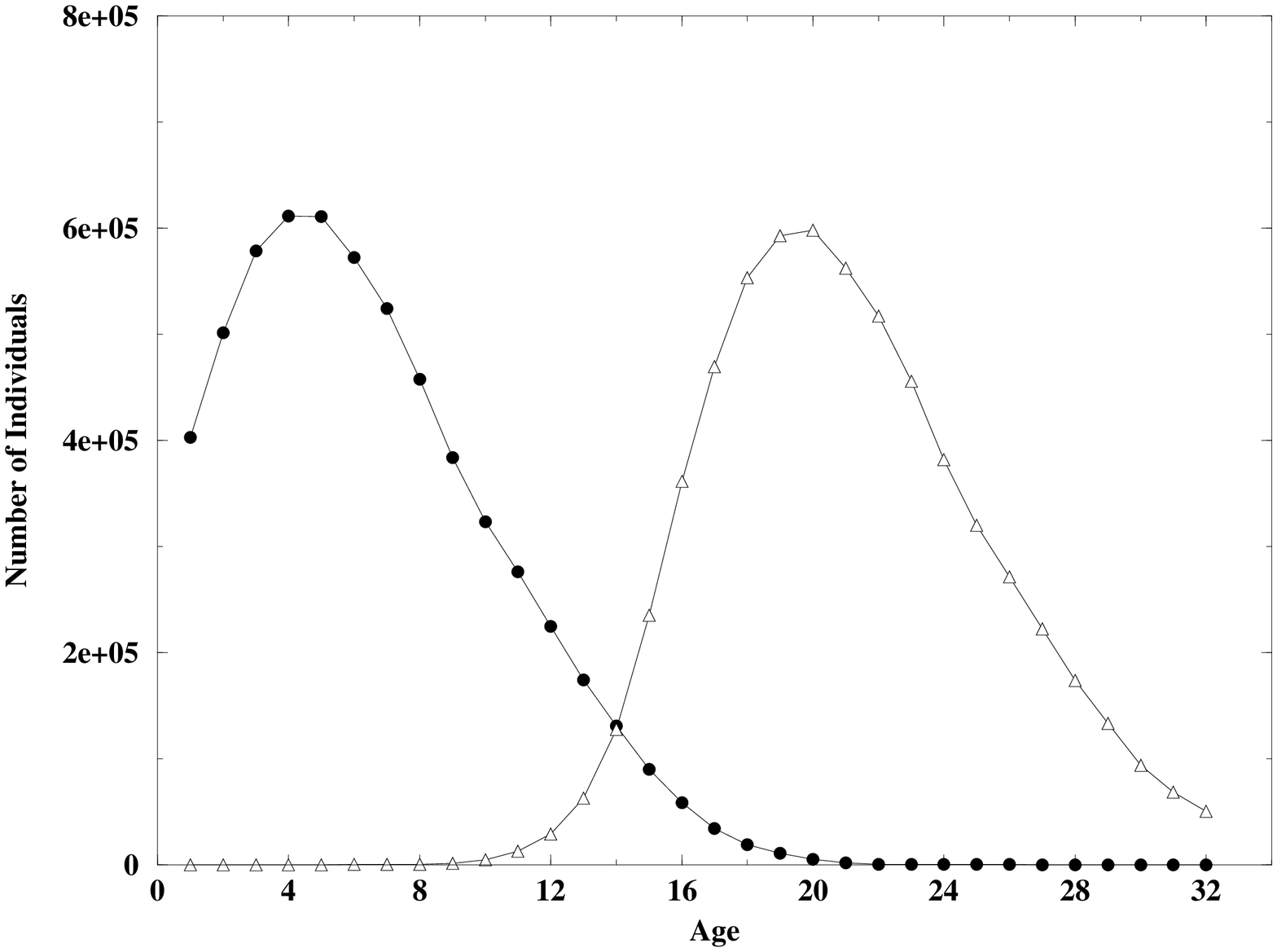}
\end{center}
\caption{
Distribution of minimum age of reproduction (circles) and of genetic
death age (triangles) after a stationary state has been reached for
the sexual population. Parameters: initial population = 25,000 males 
plus 25,000 females; maximum population size $N_{max} = 500,000$, 
initial $a_m = 1$ and initial $a_d = 16$. 
} 
\end{figure}

As in many previous simulations, starting with Redfield
\cite{redfield,anais}, 
the sexual version does not give a clear advantage over the asexual one and
additional assumptions like in \cite{pleio} may be needed to justify sex.

\section{Conclusion}

The simple alternative \cite{dresden} to the more widespread Penna model 
\cite{penna,book} of biological ageing via inherited mutations was modified here
to give reasonable agreement with the Gompertz mortality law at middle and
old age, to allow a more realistic description of spatial fluctuations on a
lattice, and to include sexual reproduction. In all cases the results were
similar to those obtained earlier in the Penna model. Further research could
concentrate on the simplification \cite{pmco}, to omit the restriction $a_d \le
32$ and to omit the parameter now taken as 0.08.
\bigskip

\noindent {\bf Acknowledgements}: A. O. Sousa and S. Moss de Oliveira
thank CAPES, CNPq and FAPERJ for financial support; DS thanks the
J\"ulich Supercomputer Center for time on their Cray-T3E.

\end{document}